\pdfoutput=1
\documentclass[11pt,twoside,a4paper,cmspaper,final,collab]{cms-tdr}
\def\svnVersion{fc1c2fa}\def\svnDate{2023/08/07}\def\cmsCernNoTag{CERN-EP-2022-081}\def\cmsCernDate{\today}\def\cmsMessage{Published in Physical Review Letters as \href{http://dx.doi.org/10.1103/PhysRevLett.131.061801}{\doi{10.1103/PhysRevLett.131.061801}.}}
\begin{document}\cmsNoteHeader{HIG-21-008}

\newcommand{\bq}{\PQb}
\newcommand{\cq}{\PQc}
\newcommand{\vhcc}{\ensuremath{{\PV\PH(\PH\to\ccbar)}}\xspace}
\newcommand{\vhbb}{\ensuremath{{\PV\PH(\PH\to\bbbar)}}\xspace}
\newcommand{\vzbb}{\ensuremath{{\PV\PZ(\PZ\to\bbbar)}}\xspace}
\newcommand{\vzcc}{\ensuremath{{\PV\PZ(\PZ\to\ccbar)}}\xspace}
\newcommand{\HCC}{\ensuremath{{\PH\to\ccbar}}\xspace}
\newcommand{\HBB}{\ensuremath{{\PH\to\bbbar}}\xspace}
\newcommand{\ZCC}{\ensuremath{{\PZ\to\ccbar}}\xspace}
\newcommand{\HTT}{\ensuremath{{\PH\to\PGt\PGt}}\xspace}
\newcommand{\pb}{\unit{pb}}
\newcommand{\ptV}{\ensuremath{\pt(\PV)}\xspace}
\newcommand{\ptH}{\ensuremath{\pt(\PH)}\xspace}
\newcommand{\larger}{large-\ensuremath{R}}
\newcommand{\smallr}{small-\ensuremath{R}}
\newcommand{\VZ}{\ensuremath{\PV\PZ}\xspace}
\newcommand{\VH}{\ensuremath{\PV\PH}\xspace}
\newcommand{\WH}{\ensuremath{\PW\PH}\xspace}
\newcommand{\ZH}{\ensuremath{\PZ\PH}\xspace}
\newcommand{\WW}{\ensuremath{\PW\PW}\xspace}
\newcommand{\WZ}{\ensuremath{\PW\PZ}\xspace}
\newcommand{\ZZ}{\ensuremath{\PZ\PZ}\xspace}
\newcommand{\kappaC}{\ensuremath{\kappa_{\PQc}}\xspace}
\newcommand{\hcand}{\ensuremath{\PH_{\text{cand}}}\xspace}
\newcommand{\mhcand}{\ensuremath{m\left(\hcand\right)}\xspace}
\newcommand{\vjets}{\PV{}+jets\xspace}
\newcommand{\sigmaVH}[1][{}]{\ensuremath{\sigma_{\text{#1}}\left(\PV\PH\right)}\xspace}
\newcommand{\brHCC}[1][{}]{\ensuremath{\mathcal{B}_{\text{#1}}\left(\PH\to\PQc\PAQc\right)}\xspace}
\newcommand{\sigmaB}[1][{}]{\ensuremath{\sigmaVH[#1]\brHCC[#1]}\xspace}
\newcommand{\muVHcc}{\ensuremath{\mu_{\vhcc}}\xspace}
\newcommand{\muVZcc}{\ensuremath{\mu_{\vzcc}}\xspace}
\newcommand{\cvsb}{\textit{C\text{vs}B}\xspace}
\newcommand{\cvsl}{\textit{C\text{vs}L}\xspace}
\ifthenelse{\boolean{cms@external}}{\providecommand{\cmsLeft}{upper\xspace}}{\providecommand{\cmsLeft}{left\xspace}}
\ifthenelse{\boolean{cms@external}}{\providecommand{\cmsRight}{lower\xspace}}{\providecommand{\cmsRight}{right\xspace}}

\cmsNoteHeader{HIG-21-008}
\title{Search for Higgs boson decay to a charm quark-antiquark pair in proton-proton collisions at \texorpdfstring{$\sqrt{s}=13\TeV$}{sqrt(s) = 13 TeV}}

\author*[cern]{CMS Collaboration}

\date{\today}

\abstract{
A search for the standard model Higgs boson decaying to a charm quark-antiquark pair, \HCC,  produced in association with a leptonically decaying \PV (\PW or \PZ) boson is presented. The search is performed with proton-proton collisions at ${\sqrt{s}=13\TeV}$ collected by the CMS experiment,  corresponding to an integrated luminosity of 138\fbinv.  Novel charm jet identification and analysis methods using machine learning techniques are employed.  The analysis is validated by searching for \ZCC in \VZ events,  leading to its first observation at a hadron collider with a significance of 5.7 standard deviations.  The observed (expected) upper limit on \sigmaB is 0.94 ($0.50^{+0.22}_{-0.15}$)\pb  at $95\%$ confidence level (\CL),  corresponding to 14 ($7.6^{+3.4}_{-2.3}$) times the standard model prediction.  For the Higgs-charm Yukawa coupling modifier, $\kappaC$,  the observed (expected) $95\%$ \CL interval is ${1.1 <\abs{\kappaC}<5.5}$ (${\abs{\kappaC}<3.4}$),  the most stringent constraint to date.
}

\hypersetup{%
pdfauthor={CMS Collaboration},%
pdftitle={Search for Higgs boson decay to a charm quark-antiquark pair in proton-proton collisions at sqrt(s) = 13 TeV},%
pdfsubject={CMS},%
pdfkeywords={CMS,  Higgs boson, charm quark, jet tagging}} %

\maketitle

The discovery of a Higgs boson (\PH) with the LHC Run 1 data by the ATLAS~\cite{Aad:2012tfa} and 
CMS~\cite{Chatrchyan:2012xdj,CMS:2013btf} experiments in 2012 
was a major advance in the understanding of the electroweak (EW) symmetry breaking mechanism. 
The measured Higgs boson mass is $125.38 \pm 0.14$\GeV~\cite{CMS:2020xrn}. 
The observed interactions with gauge bosons and third-generation fermions~\cite{Aad:2014eha,Khachatryan:1728107,Aad:2014eva,Chatrchyan:2013mxa,ATLAS:2014aga,Aad:2015ona,Chatrchyan:1633401,Aad:2015vsa,Chatrchyan:1643937,Sirunyan:2273798,Sirunyan:2018egh}, 
and all measured properties~\cite{Aad:2015gba,Khachatryan:2014jba,Chatrchyan:2012jja,Aad:2013xqa,Khachatryan:2016vau,Aad:2015zhl,Sirunyan:2017exp,Aaboud:2018ezd,Aaboud:2018wps,CMS:2020xrn}, 
are compatible with standard model (SM) predictions. 
Recently, the CMS Collaboration reported the first evidence of Higgs boson decays to muons, \ie, second-generation leptons~\cite{Sirunyan:2018hbu}. 
An important next milestone is the observation of its coupling to second-generation quarks.
In this Letter, we focus on a search for Higgs boson decay to $\ccbar$, 
a charm quark-antiquark pair. 
The corresponding Yukawa coupling, $y_{\cq}$, can be significantly modified in the 
presence of physics beyond the SM~\cite{Delaunay:2013pja,Perez:2015aoa,Ghosh:2015gpa,Botella:2016krk,Harnik:2012pb,Altmannshofer:2016zrn}. 
However, the small branching fraction predicted by the SM, 
ubiquitous production of quark and gluon jets at the LHC, 
and the difficulty of identifying charm quark jets in a hadronic environment, 
including distinguishing them from bottom quark jets,
make this a challenging measurement.
Searches for \HCC 
reported in Refs.~\cite{Aaboud:2018fhh,Sirunyan:2019qia,ATLAS:2022ers} by the ATLAS and CMS Collaborations
target the associated production of a Higgs boson with a \PV (\PW or \PZ) boson.
Using 139\fbinv\ of data at 13\TeV, 
the most recent search by the ATLAS Collaboration obtains an observed (expected) upper limit on the product of the production cross section \sigmaVH
and branching fraction \brHCC of 26 (31) times the SM prediction at 95\% confidence level (\CL)~\cite{ATLAS:2022ers}.

This Letter presents a search for \HCC in \VH production using proton-proton ($\Pp\Pp$) collision data at $\sqrt{s}=13\TeV$,
collected by the CMS detector in 2016--2018, 
and corresponding to an integrated luminosity of 138\fbinv~\cite{CMS-LUM-17-003,CMS-PAS-LUM-17-004,CMS-PAS-LUM-18-002}.
Building upon a previous analysis by the CMS Collaboration~\cite{Sirunyan:2019qia},
advanced \PQc jet reconstruction and identification algorithms and sophisticated analysis techniques
using machine learning (ML) have been further developed to significantly improve sensitivity.

The CMS apparatus~\cite{CMS:2008xjf} is a multipurpose, nearly hermetic
detector, designed to trigger on~\cite{CMS:2020cmk,CMS:2016ngn} and identify
electrons, muons, photons, and (charged and neutral)
hadrons~\cite{CMS:2020uim,CMS:2018rym,CMS:2014pgm}. A global
reconstruction ``particle-flow" (PF) algorithm~\cite{CMS:2017yfk} combines the
information provided by the all-silicon inner tracker and by the crystal
electromagnetic and brass-scintillator hadron calorimeters, operating inside a
3.8\unit{T} superconducting solenoid, with data from gas-ionization muon
detectors embedded in the solenoid flux-return yoke, 
to reconstruct charged and neutral hadrons, electrons, photons, 
and muons (PF candidates) and build \PGt leptons,
jets, missing transverse momentum, and other physics
objects~\cite{CMS:2018jrd,CMS:2016lmd,CMS:2019ctu}.

Two collections of jets, formed by PF candidates clustered with the anti-\kt 
algorithm~\cite{Cacciari:2008gp, Cacciari:2011ma}, are used in the search. 
The first uses a distance parameter $R=0.4$, 
and will be referred to as ``\smallr'' jets. 
The second uses $R=1.5$ and contains what are referred to as ``\larger'' jets. 
The impact of particles from additional $\Pp\Pp$ interactions within the same or nearby bunch crossings (pileup)
is mitigated via the charged hadron subtraction algorithm~\cite{CMS:2017yfk} for \smallr\ jets,
and via the PUPPI~\cite{CMS:2020ebo,Bertolini:2014bba} algorithm for \larger\ jets.
A regression algorithm~\cite{CMS-DP-2021-017} is developed to improve \larger\ jet mass reconstruction 
that exploits properties of the PF candidates and secondary vertices associated to the jet 
using the \textsc{ParticleNet} graph neural network~\cite{Qu:2019gqs}.  
Mass resolution is improved by about 50\% over 
traditional jet grooming algorithms~\cite{Dasgupta:2013ihk,Butterworth:2008iy}.
The \smallr\ (\larger) jets are required to have transverse momentum (\pt) above 25\,(200)\GeV and 
to be within the tracker acceptance.

Signal and background processes are simulated using various Monte Carlo event
generators. The detector response is modeled with
\GEANTfour~\cite{GEANT4}. The \WH and quark-induced \ZH signal
processes are generated at next-to-leading order (NLO) accuracy in quantum chromodynamics (QCD)
using the {\POWHEG} v2~\cite{Nason:2004rx,POWHEG,Alioli:2010xd} event
generator extended with the multi-scale improved NLO (MiNLO)
procedure~\cite{Hamilton2012,Luisoni:2013kna}, while the gluon-induced
\ZH process is generated at leading order (LO) accuracy with {\POWHEG} v2.
The Higgs boson mass is set to 125\GeV for all simulations. 
The production cross sections of the signal processes~\cite{deFlorian:2016spz} 
are corrected as a function of \ptV to next-to-next-to-leading order (NNLO) QCD + NLO EW
accuracy combining the \textsc{vhnnlo}~\cite{Ferrera:2013yga,Ferrera:2014lca,Ferrera:2011bk,Ferrera:2017zex},
\textsc{vh@nnlo}~\cite{Brein:2012ne,Harlander:2013mla}, 
and \textsc{hawk} v2.0~\cite{Denner:2014cla} generators, as described in Ref.~\cite{deFlorian:2016spz}.

The \vjets background samples 
are generated with {\MGvATNLO} v2.6.0~\cite{Alwall:2014hca} at 
NLO with up to two additional partons. 
The top quark pair (\ttbar)~\cite{Frixione:2007nw} and single top quark production processes
\cite{Re:2010bp,Frederix:2012dh,Alioli:2009je}
are generated to NLO accuracy with {\POWHEG} v2.
The production cross sections for the \ttbar samples are scaled to the
NNLO prediction with the next-to-next-to-leading logarithmic resummation result obtained from {{\TOPpp v2.0}}~\cite{Czakon:2011xx},
and the differential cross sections as a function of top quark \pt are corrected 
to the NNLO QCD + NLO EW prediction~\cite{Czakon:2017wor}.
Diboson backgrounds are generated at NLO with {\POWHEG} v2 ({\MGvATNLO} v2.4.2) 
for the \WW~\cite{Melia:2011tj} (\WZ and \ZZ) process. 
Production cross sections of the diboson processes are reweighted 
as a function of the subleading vector boson \pt 
to NNLO QCD + NLO EW accuracy~\cite{Kallweit:2019zez}.

The NLO {NNPDF3.0}~\cite{Ball:2014uwa} (NNLO {NNPDF3.1}~\cite{Ball:2017nwa}) 
parton distribution function (PDF) set is used for the 2016 (2017 and 2018) simulations. 
For parton showering and hadronization, including the \HCC decay, 
matrix element generators are interfaced with {\PYTHIA} 
v8.230~\cite{Sjostrand:2014zea} with the 
CUETP8M1~\cite{Khachatryan:2015pea} (CP5~\cite{Sirunyan:2019dfx}) underlying event tune
for 2016 (2017 and 2018) samples.
The matching of jets from matrix element calculations and those from parton showers
is done with the FxFx~\cite{Frederix:2012ps} (MLM~\cite{Alwall:2007fs}) prescription for NLO (LO) samples.
For all samples, pileup interactions are simulated with {\PYTHIA} and added to the hard-scattering
process. Events are then reweighted to match the pileup profile observed in data.

The analysis is carried out in mutually exclusive channels targeting leptonic decays of the vector bosons: 
$\PZ\to \nu\nu$,  $\PW\to \ell \nu$, and $\PZ\to \ell\ell$, where $\ell$ is an electron or a muon, 
and referred to as the 0L, 1L, and 2L channels. 
Events are collected using triggers based on large missing transverse momentum, or the presence of one, 
or two well-identified and isolated leptons.
The event selection criteria are detailed in Ref.~\cite{Sirunyan:2019qia}.

As for the previous search~\cite{Sirunyan:2019qia}, the analysis reconstructs 
the Higgs boson candidate (\hcand) assuming either a 
``merged-jet'' topology, in which the hadronization products 
of the two charm quarks are reconstructed as a single \larger\ jet,
or a ``resolved-jet'' topology, in which the \hcand 
is reconstructed from two well-separated and individually resolved \smallr\ \cq\ jets.
These two topologies can have significant overlap and 
are made distinct in reference to whether 
a given \hcand, identified through a \larger\ jet in the event, 
has \pt above or below a threshold of 300\GeV for the definition of the merged-jet and the resolved-jet topology, respectively,
where the dividing line is chosen to maximize sensitivity to \vhcc.

On average, the \hcand \larger\ jets in the signal process have larger \pt than those from the \vjets and \ttbar backgrounds.
Thus, the high-\pt regime explored in the merged-jet topology, although amounting to less than 
5\% of the signal cross section, provides significant sensitivity to the search. 
Because the highly boosted Higgs boson decay produces a more narrowly concentrated collection of final-state particles,
the use of a single \larger\ jet enhances the probability of correctly capturing the hadronization products of both charm quarks,  
and facilitates the accounting of final-state radiation (FSR) emitted by the quarks. 
A detailed discussion of the advantages can be found in Refs.~\cite{Butterworth:2008iy,Plehn:2009rk,Sirunyan:2019qia}.

\begin{figure}[tbp]
\centering
\includegraphics[width=0.45\textwidth]{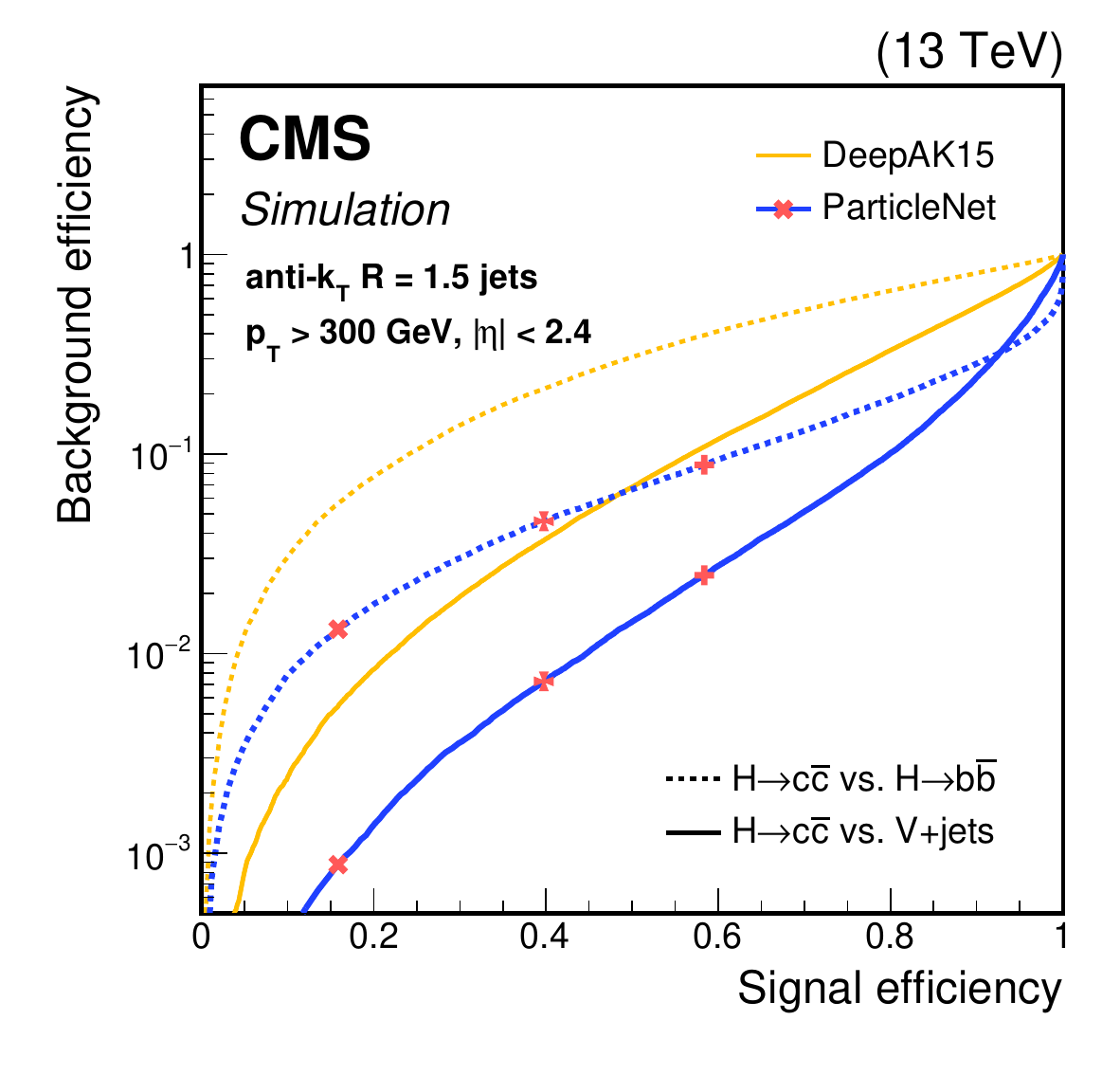}
\caption{\label{fig:roc_pnet} Performance of \textsc{ParticleNet} (blue lines) for identifying a $\ccbar$ pair 
for \larger\ jets with $\pt>300$\GeV. 
The solid (dashed) line shows the efficiency to correctly identify \HCC vs. 
the efficiency of misidentifying quarks or gluons from the \vjets process (vs. \HBB).
The red crosses represent the three working points used in the merged-jet analysis.
The performance of \textsc{DeepAK15} (yellow lines) used in Ref.~\cite{Sirunyan:2019qia}
is shown for comparison.}
\end{figure}

State-of-the-art performance in the reconstruction and identification of the pair of \cq\ quarks 
from the Higgs boson decay is achieved with \textsc{ParticleNet}~\cite{Qu:2019gqs}, a novel jet identification algorithm. 
Using PF candidates and secondary vertices associated to \larger\ jets as inputs, 
\textsc{ParticleNet} simultaneously exploits information related to jet substructure, flavor, and pileup  
with a graph neural network~\cite{Wang:2018nkf},
yielding substantial gains over other approaches~\cite{CMS-DP-2020-002,Kasieczka:2019dbj}. 
Decorrelation of the algorithm's response with the jet mass is achieved by training it
with a dedicated set of simulations produced with the same jet mass distributions for the signal and background 
processes~\cite{CMS-DP-2020-002}. 
Figure~\ref{fig:roc_pnet} shows the performance of the $\ccbar$
discriminant in identifying a pair of \cq quarks from Higgs boson decay for \larger\ jets 
with $\pt>300$\GeV. 
\textsc{ParticleNet} is compared to the previous state-of-the-art 
$\ccbar$ discriminant ``\textsc{DeepAK15}''~\cite{Sirunyan:2019qia,CMS:2020poo}, yielding
an improvement by a factor of 4 to 7 in the rejection of other jet flavors. 
Three working points are defined on the $\ccbar$ discriminant distribution 
with approximately 58, 40, and 16\% efficiencies for identifying a $\ccbar$ pair. 
The corresponding misidentification rates of light quark and gluon jets ($\bbbar$ jets) 
are 2\,(9), 0.7\,(5), and 0.08\,(1)\%.
These working points are used to separate events into three mutually exclusive 
categories with different $\ccbar$ purity to improve the sensitivity of the analysis. 
The $\ccbar$ identification efficiency in data is measured using a sample of events 
containing a gluon splitting to $\ccbar$. 
To increase the similarity to \HCC decay,
a dedicated BDT classifier is developed to enrich jets where a large fraction of momentum 
is carried by the quark pair from gluon splitting 
rather than by additionally radiated gluons~\cite{CMS-DP-2022-005}.
The \pt-dependent data-to-simulation efficiency ratios (used as corrective scale factors) 
are typically 0.9--1.3 with corresponding uncertainties of 20--30\%.

The main backgrounds, \ttbar and \vjets, are suppressed by a separate
boosted decision tree (BDT) classifier for each channel,
using kinematical variables 
that are not correlated with the \hcand mass, \mhcand, or the $\ccbar$ discriminant as inputs. 
The BDT design relies on previous developments~\cite{Sirunyan:2019qia}
with improvements in variable selection and training procedure, 
leading to ${\approx}$15\% enhancement of the sensitivity of the analysis. 
The BDT discriminants are used to define 2 (1) signal regions (SRs) in the 1L (0L and 2L) channel.
Events in the SRs are further subdivided into the three $\ccbar$ discriminant categories mentioned above.
The \mhcand distributions are used to separate signal and background contributions in each SR, 
as both the BDTs and the $\ccbar$ discriminant are designed to be largely independent of \mhcand.

More than 95\% of the \VH events have a Higgs boson with $\ptH<300\GeV$, 
corresponding to the phase space region where
the Higgs boson decay products generally give rise to two distinctly reconstructed \smallr\ jets.
The resolved-jet topology exploits a large fraction of this phase space, which, 
however, contains higher background contamination than that used in the merged-jet analysis.
The \hcand\ is reconstructed via two distinct \smallr\ jets.
The identification of \cq jets relies on 
the ML-based \textsc{DeepJet} algorithm~\cite{Bols:2020bkb,CMS-DP-2018-058}. 
The discrimination between \cq jets and light quark or gluon jets (\bq jets) is achieved via 
the ratio of the corresponding probabilities, defined as
\cvsl (\cvsb)~\cite{CMS:2021scf}.
Potential differences in the discriminant shapes between data and simulation are 
taken into account by flavor-dependent simulation-to-data efficiency scale factors~\cite{CMS:2021scf},
which are typically in the 0.9--1.0 range.
The corresponding uncertainties range from ${\approx}$2\% for light quark and gluon jets or \bq\ jets,
to ${\approx}$5\% for \PQc jets.

The two jets with the highest \cvsl discriminants in each event are selected
to reconstruct the \hcand\ momentum four-vector. 
To improve sample purity, criteria are imposed on \cvsl and \cvsb 
for the leading jet that correspond to ${\approx}40\%$ \cq jet efficiency, 
for ${\approx}4$\,(${\approx}16$)\% light quark or gluon jet (\bq jet) misidentification rate.
Multiple corrections are applied to improve the \mhcand reconstruction. 
To account for a potential underestimation of the \cq jet energy, 
due to the presence of undetected neutrinos in \cq\ hadron decays, 
an ML-based jet energy regression algorithm~\cite{CMS:2019uxx}, originally developed for \bq jets, is adapted for \cq jets.
In addition, \smallr\ jets reconstructed in the vicinity of \hcand\ jets,
often stemming from FSR, are included in the \hcand\ reconstruction~\cite{Sirunyan:2019qia}.
The \mhcand resolution in the 2L channel is further improved 
via a kinematic fit by balancing the momenta of the two \smallr\ jets 
and the lepton pair within experimental uncertainties~\cite{Sirunyan:2018kst}. 
These steps improve the \mhcand\ resolution up to 20\%.
Finally, a BDT classifier is developed to maximize the discrimination power between signal 
and background processes in each channel, using event-level kinematical variables, 
\PQc~jet identification discriminants, and properties of \hcand, including \mhcand, as inputs~\cite{Sirunyan:2019qia}.

The signal strength modifier $\mu$, defined as 
${\left(\sigma\mathcal{B}\right)_{\text{obs}}/\left(\sigma\mathcal{B}\right)_{\text{SM}}}$ 
where $\sigma$ is the signal production cross section and $\mathcal{B}$ is the branching fraction,
is measured via a binned maximum likelihood fit to data.
The best-fit value of $\mu$ and an approximate 68\% \CL confidence interval 
are extracted following the procedure in Ref.~\cite{Khachatryan:2014jba}.
The fitted variable is \mhcand in the merged-jet analysis, 
and the BDT discriminant in the resolved-jet analysis.
The normalizations of the main backgrounds, namely \vjets and \ttbar, 
are estimated by including dedicated control regions in the fit 
following the strategy detailed in Ref.~\cite{Sirunyan:2019qia}. 
Contributions from single top, diboson and \vhbb\ processes are estimated from simulation assuming SM production rates.
Because of improvements in \PQc~jet identification, 
the difficult \vhbb\ background contribution in the high $\ccbar$ purity SRs is 
reduced from about twenty times to only about twice that expected for \vhcc, 
and the \vzbb yield is reduced from more than 100\% to about 10\% of that expected for \vzcc.
Contributions from \HTT decays are negligible after the above-mentioned selection criteria, 
and are not considered.

Systematic uncertainties affecting normalizations
and shapes of fitted variables are taken into account via nuisance parameters.
The relative contribution of each uncertainty source to the total uncertainty 
in the fitted $\mu$ is summarized in Table~\ref{tab:syst-breakdown}.
The leading uncertainty is statistical because of the limited number of events 
in the SRs as well as the control regions used to extract background normalizations.
The main experimental systematic uncertainties are associated with limited simulation sample sizes, 
particularly due to large fractions of negatively weighted events in the NLO \vjets samples,  
and the \PQc~jet identification efficiencies, representing ${\approx}37$ and ${\approx}23\%$ 
of the total, respectively. 
Theoretical uncertainties in the cross sections, \pt spectra, PDFs, renormalization and
factorization scales, represent ${\approx}22\%$ of the total uncertainty in $\mu$.

\begin{table}[tbp]
\topcaption{The relative contributions to the total uncertainty in the signal strength modifier $\mu$ for the \vhcc process, 
where the best fit is ${\muVHcc = 7.7^{+3.8}_{-3.5}}$.}
\label{tab:syst-breakdown}
\centering
{
\begin{scotch}{lr}
Uncertainty source                                      &   $\Delta\mu/\left(\Delta\mu\right)_{\text{tot}}$ \\
\hline

{Statistical}                                    &   {85\%} \\
\hspace{2em}    Background normalizations               &   37\%          \\

{Experimental}                                   &  {48\%} \\
\hspace{2em}    Sizes of the simulated samples          &   37\%          \\
\hspace{2em}    \PQc~jet identification efficiencies    &   23\%          \\
\hspace{2em}    Jet energy scale and resolution         &   15\%          \\
\hspace{2em}    Simulation modeling                     &   11\%          \\
\hspace{2em}    Integrated luminosity                   &   6\%           \\
\hspace{2em}    Lepton identification efficiencies      &   4\%           \\

{Theory}                                         &   {22\%} \\
\hspace{2em}    Backgrounds                             &   17\%          \\
\hspace{2em}    Signal                                  &   15\%          \\

\end{scotch}
}
\end{table}

\begin{figure}[tbp]
\centering
\includegraphics[width=0.45\textwidth]{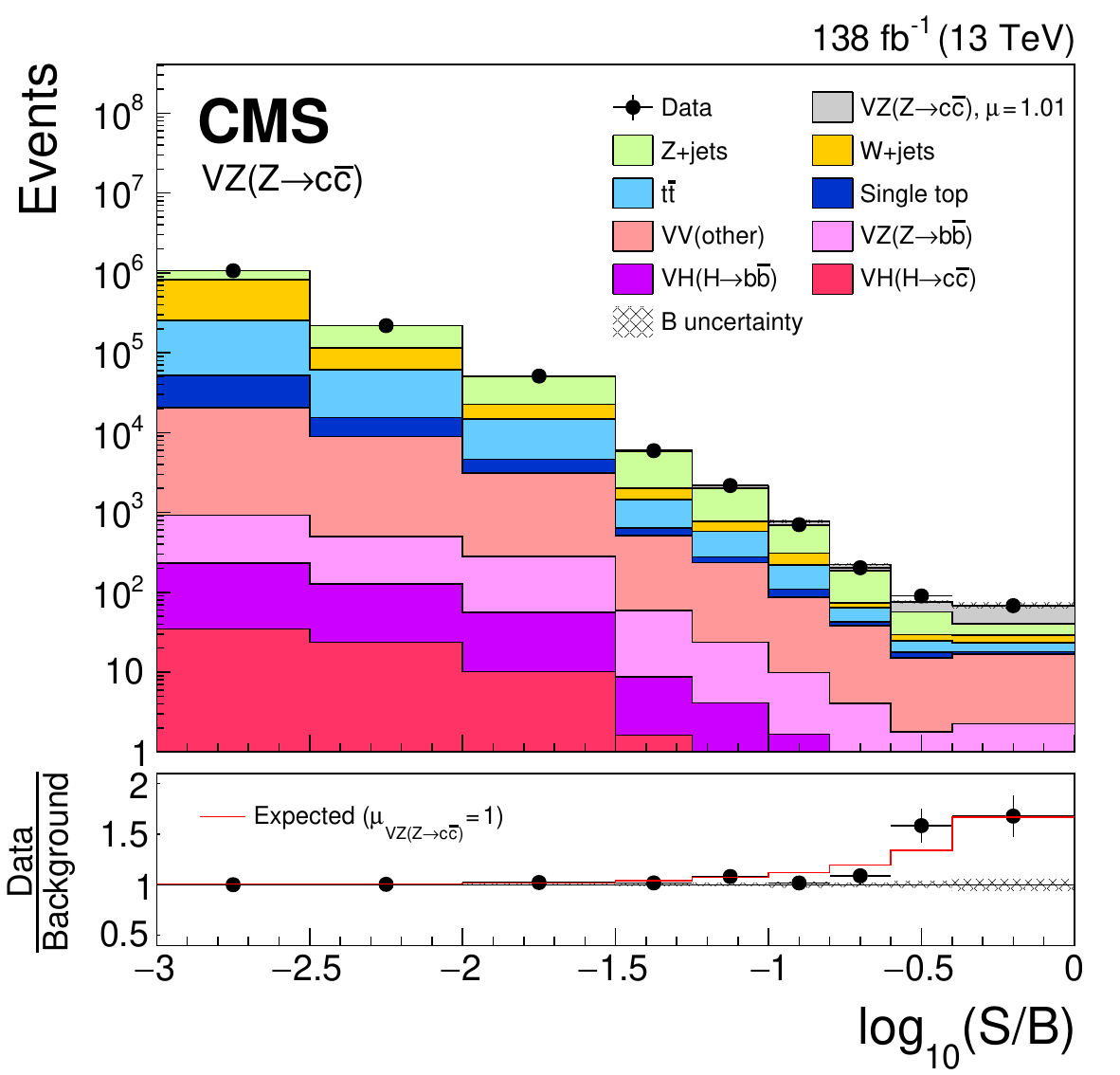}
\includegraphics[width=0.45\textwidth]{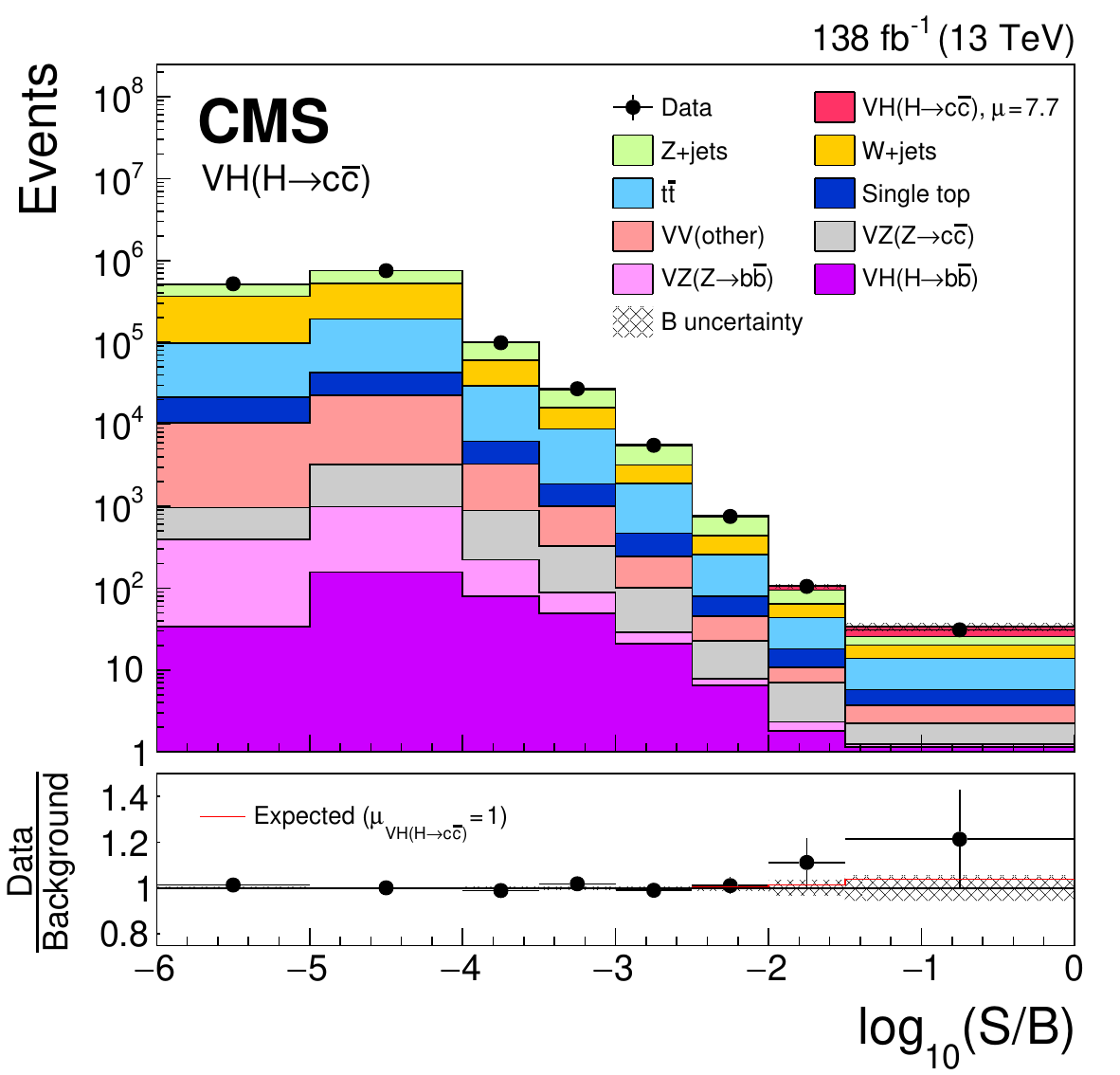}
\caption{Distribution of events as a function of $S/B$
in the \vzcc (\cmsLeft) and \vhcc (\cmsRight) searches, where $S$ and $B$ are the 
postfit signal and background yields, respectively, 
in each bin of the fitted \mhcand or BDT discriminant distributions. 
The bottom panel shows the ratio of data to the total background, 
with the uncertainty in background indicated by gray hatching. 
The red line represents background plus SM signal divided by background.}
\label{fig:s_over_b_ordered}
\end{figure}

A search for the analogous SM process \vzcc is performed to validate the analysis strategy. 
The BDTs in the resolved-jet topology are modified by training them 
with \vzcc as signal.
No modification is needed for the BDTs in the merged-jet topology as they are independent of \mhcand. 
The best fit $\mu$ of this process is ${\muVZcc=1.01_{-0.21}^{+0.23}}$, 
in agreement with the SM expectation. 
Figure~\ref{fig:s_over_b_ordered}~(\cmsLeft) shows the distribution of events 
in all channels, sorted into bins of similar signal-to-background ratios. 
The observed data show a visible excess over the expected backgrounds.
The significance of the excess is computed using the asymptotic distribution of a test statistic based on 
the profile likelihood ratio~\cite{CMS-NOTE-2011-005,Cowan:2010js}.
The observed (expected) significance is 4.4 (4.7) standard deviations 
for the merged-jet analysis,
3.1 (3.3) standard deviations for the resolved-jet analysis, 
and 5.7 (5.9) standard deviations for their combination.
This is the first observation of \ZCC\ at a hadron collider.

\begin{figure}[tbp]
\centering
\includegraphics[width=0.45\textwidth]{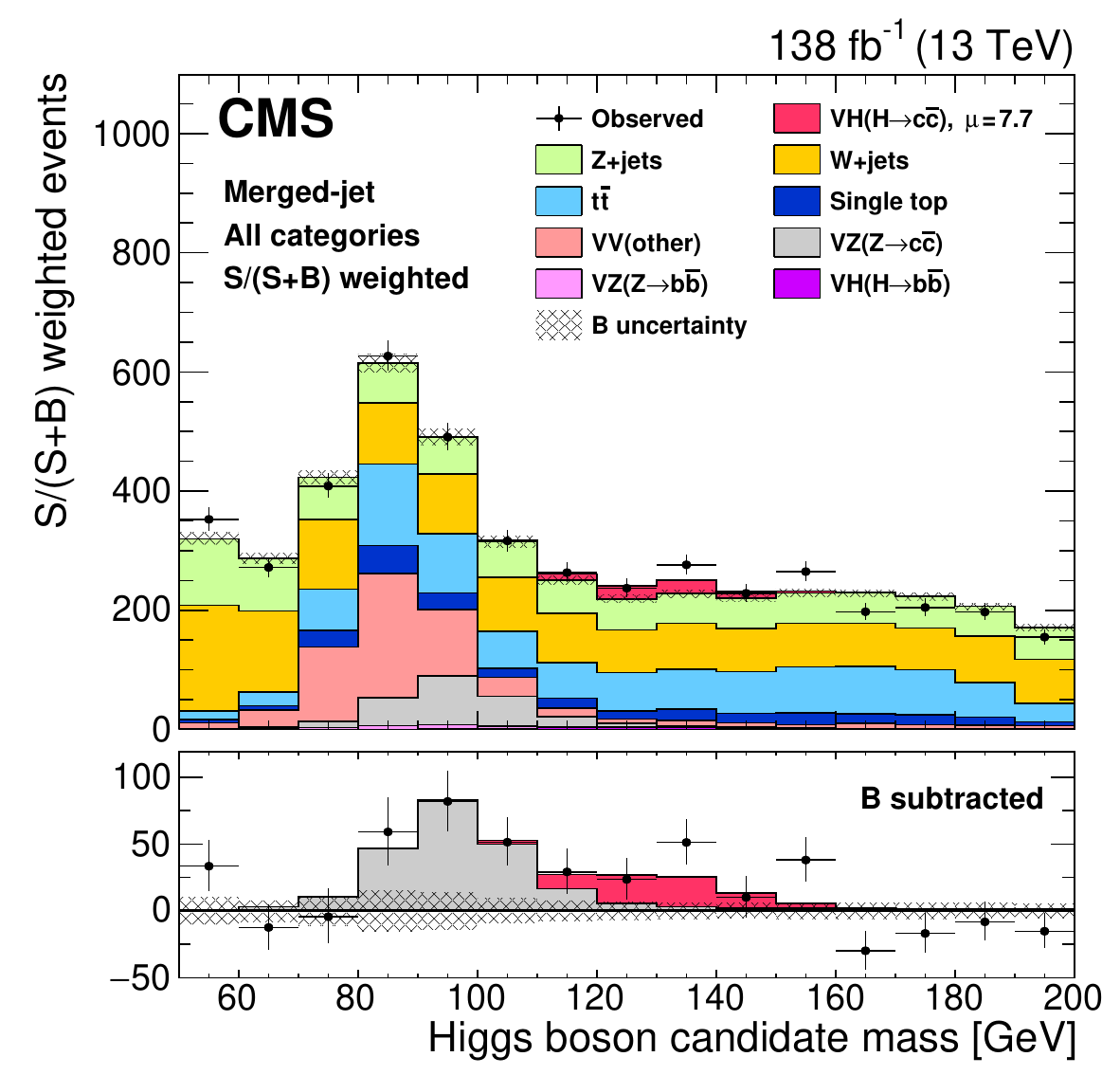}
\caption{Combined \mhcand distribution in all channels of the merged-jet analysis. 
The fitted \mhcand distribution in each SR is weighted by $S/(S+B)$, where $S$ and $B$ are 
the postfit \vhcc signal and total background yields.
The lower panel shows data (points) and the fitted \vhcc (red) and \vzcc (grey) distributions 
after subtracting all other processes. Error bars represent 
pre-subtraction statistical uncertainties in data, while the gray hatching indicates 
the total uncertainty in the signal and all background processes.}
\label{fig:mass_merged_jet}
\end{figure}

\begin{figure}[tbh!]
\centering
\includegraphics[width=0.45\textwidth]{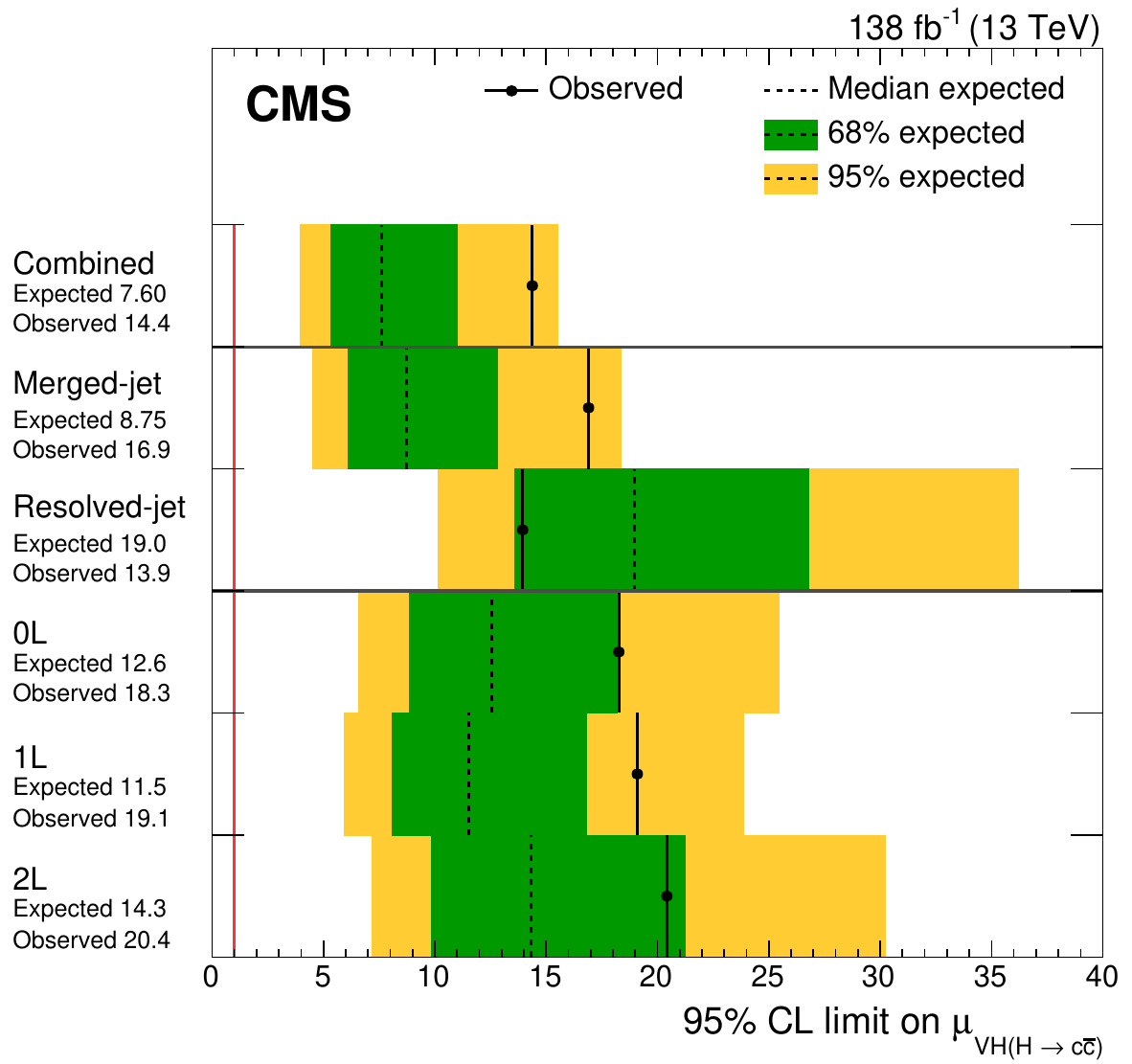}
\caption{The 95\% \CL upper limits on \muVHcc. 
Green and yellow bands indicate the 68 and 95\% intervals on the expected limits, respectively. 
The vertical red line indicates the SM value $\muVHcc=1$.}
\label{fig:limits}
\end{figure}

Figure~\ref{fig:s_over_b_ordered}~(\cmsRight) compares the observed data to the SM prediction in the search for \vhcc,
where the best fit is ${\muVHcc = 7.7^{+3.8}_{-3.5}}$. 
The fitted \mhcand distribution in the merged-jet topology is displayed in Fig.~\ref{fig:mass_merged_jet}.
No significant excess over the background-only hypothesis is observed. 
An upper limit on \muVHcc is extracted using the \CLs criterion~\cite{Junk:1999kv,Read:2002hq}. 
The test statistic is the profile likelihood ratio modified for upper limits~\cite{CMS-NOTE-2011-005}, 
and the asymptotic approximation~\cite{Cowan:2010js} is used in the limit setting procedure.
The observed (expected) 95\% \CL upper limit on \muVHcc is 14 ($7.6^{+3.4}_{-2.3}$), 
which is equivalent to an observed (expected) upper limit on \sigmaB
of 0.94 ($0.50^{+0.22}_{-0.15}$)\pb. 
Contributions from the individual channels are summarized in Fig.~\ref{fig:limits}. 
Tabulated results are provided in the HEPData record for this analysis~\cite{hepdata}.

The result is interpreted in the $\kappa$-framework~\cite{Heinemeyer:2013tqa,deFlorian:2016spz} 
by reparameterizing \muVHcc in terms of the Higgs-charm Yukawa coupling modifier \kappaC, 
assuming only the Higgs boson decay widths are altered:
\begin{equation}
    \mu_{\vhcc} = \frac{\kappaC^2}{1+\mathcal{B}_{\text{SM}}\left(\HCC \right)\,(\kappaC^2-1)}.
\end{equation}
The observed 95\% \CL interval is ${1.1 < \abs{\kappaC} < 5.5}$, 
and the corresponding expected constraint is ${\abs{\kappaC} < 3.4}$.

In summary, a search for the SM Higgs boson 
decaying to a pair of charm quarks in the CMS experiment is presented.
Novel jet reconstruction and identification tools, and analysis techniques are 
developed for this analysis, which is validated by measuring the \vzcc process. 
The observed \PZ boson signal relative to the SM prediction is ${\muVZcc = 1.01_{-0.21}^{+0.23}}$, 
with an observed (expected) significance of 5.7 (5.9) standard deviations above the background-only hypothesis.
This is the first observation of \ZCC\ at a hadronic collider.

The observed (expected) upper limit on \sigmaB is 0.94 ($0.50^{+0.22}_{-0.15}$)\pb, 
corresponding to 14 ($7.6^{+3.4}_{-2.3}$) times the theoretical prediction 
for an SM Higgs boson mass of 125.38\GeV.
The observed (expected) 95\% \CL interval on the modifier, \kappaC, 
for the Yukawa coupling of the Higgs boson to the charm quark 
is ${1.1 < \abs{\kappaC} < 5.5}$ (${\abs{\kappaC} < 3.4}$). 
This is the most stringent constraint on \kappaC to date.

\begin{acknowledgments}
We congratulate our colleagues in the CERN accelerator departments for the excellent performance of the LHC and thank the technical and administrative staffs at CERN and at other CMS institutes for their contributions to the success of the CMS effort. In addition, we gratefully acknowledge the computing centres and personnel of the Worldwide LHC Computing Grid and other centres for delivering so effectively the computing infrastructure essential to our analyses. Finally, we acknowledge the enduring support for the construction and operation of the LHC, the CMS detector, and the supporting computing infrastructure provided by the following funding agencies: BMBWF and FWF (Austria); FNRS and FWO (Belgium); CNPq, CAPES, FAPERJ, FAPERGS, and FAPESP (Brazil); MES and BNSF (Bulgaria); CERN; CAS, MoST, and NSFC (China); MINCIENCIAS (Colombia); MSES and CSF (Croatia); RIF (Cyprus); SENESCYT (Ecuador); MoER, ERC PUT and ERDF (Estonia); Academy of Finland, MEC, and HIP (Finland); CEA and CNRS/IN2P3 (France); BMBF, DFG, and HGF (Germany); GSRI (Greece); NKFIH (Hungary); DAE and DST (India); IPM (Iran); SFI (Ireland); INFN (Italy); MSIP and NRF (Republic of Korea); MES (Latvia); LAS (Lithuania); MOE and UM (Malaysia); BUAP, CINVESTAV, CONACYT, LNS, SEP, and UASLP-FAI (Mexico); MOS (Montenegro); MBIE (New Zealand); PAEC (Pakistan); MES and NSC (Poland); FCT (Portugal); MESTD (Serbia); MCIN/AEI and PCTI (Spain); MOSTR (Sri Lanka); Swiss Funding Agencies (Switzerland); MST (Taipei); MHESI and NSTDA (Thailand); TUBITAK and TENMAK (Turkey); NASU (Ukraine); STFC (United Kingdom); DOE and NSF (USA).    
\end{acknowledgments}

 \bibliography{auto_generated}
       
\cleardoublepage \appendix\section{The CMS Collaboration \label{app:collab}}\begin{sloppypar}\hyphenpenalty=5000\widowpenalty=500\clubpenalty=5000
\cmsinstitute{Yerevan Physics Institute, Yerevan, Armenia}
{\tolerance=6000
A.~Tumasyan\cmsAuthorMark{1}\cmsorcid{0009-0000-0684-6742}
\par}
\cmsinstitute{Institut f\"{u}r Hochenergiephysik, Vienna, Austria}
{\tolerance=6000
W.~Adam\cmsorcid{0000-0001-9099-4341}, J.W.~Andrejkovic, T.~Bergauer\cmsorcid{0000-0002-5786-0293}, S.~Chatterjee\cmsorcid{0000-0003-2660-0349}, K.~Damanakis\cmsorcid{0000-0001-5389-2872}, M.~Dragicevic\cmsorcid{0000-0003-1967-6783}, A.~Escalante~Del~Valle\cmsorcid{0000-0002-9702-6359}, P.S.~Hussain\cmsorcid{0000-0002-4825-5278}, M.~Jeitler\cmsAuthorMark{2}\cmsorcid{0000-0002-5141-9560}, N.~Krammer\cmsorcid{0000-0002-0548-0985}, L.~Lechner\cmsorcid{0000-0002-3065-1141}, D.~Liko\cmsorcid{0000-0002-3380-473X}, I.~Mikulec\cmsorcid{0000-0003-0385-2746}, P.~Paulitsch, F.M.~Pitters, J.~Schieck\cmsAuthorMark{2}\cmsorcid{0000-0002-1058-8093}, R.~Sch\"{o}fbeck\cmsorcid{0000-0002-2332-8784}, D.~Schwarz\cmsorcid{0000-0002-3821-7331}, S.~Templ\cmsorcid{0000-0003-3137-5692}, W.~Waltenberger\cmsorcid{0000-0002-6215-7228}, C.-E.~Wulz\cmsAuthorMark{2}\cmsorcid{0000-0001-9226-5812}
% ...
\par}
\vskip\cmsinstskip
\dag:~Deceased\\
$^{1}$Also at Yerevan State University, Yerevan, Armenia\\
$^{2}$Also at TU Wien, Vienna, Austria\\
% ...
\end{sloppypar}
\end{document}